\documentclass{webofc} 
\usepackage{mathtools}{\tiny }
\usepackage{slashed}
\usepackage[compat=1.0.0]{tikz-feynman}
\usepackage[utf8]{inputenc}
\usepackage[export]{adjustbox}
\usepackage{wrapfig}
\usepackage{mathtools}
\usepackage{ragged2e}
\usepackage{graphicx}
\usepackage{subcaption}
\usepackage{slashed}
\usepackage{amssymb}
\usepackage{pgfplots}
\pgfplotsset{compat=1.15}
\usepackage{mathrsfs}
\usepackage{comment}
\usepackage{caption}
\usepackage{xcolor}

\usepackage{physics}
\usepackage[compat=1.0.0]{tikz-feynman}
\usepackage[utf8]{inputenc}
\usepackage[export]{adjustbox}
\usepackage{wrapfig}
\graphicspath{
    {.} 
    {images/}
    {img/}
    {files/pictures/}
    {figures/}
    {plots/
}
}

\begin{document}
\title{$N^* (1520)$ transition form factors from dispersion theory}
%
%

\author{\centering Di An\thanks{\email{di.an@physics.uu.se}} 
}

\institute{Institutionen f\"or fysik och astronomi, Uppsala universitet, Box 516, S-75120 Uppsala, Sweden}

\abstract{The electromagnetic transition form factors (TFFs) of the nucleon provide important information on the internal structure of hadrons. A model-independent dispersive calculation of the electromagnetic TFFs $N^{*}(1520)\to N$ at low energies is presented. Taking pion rescattering into consideration, we derived dispersive relations for the $N^{*}(1520)\to N$ TFFs that relate space-like and time-like regions from the first principles. Based on the space-like data from JLab and hadronic data measured by HADES, we make predictions for TFFs in the time-like region. Our predictions can be tested in future experiments (e.g. HADES).%
}
\maketitle
\section{Introduction}
\label{intro}
The electromagnetic form factors (FF) and transition form factors (TFFs) provide important information on the internal structure of nucleons and their excited states.
At very high energies, perturbative QCD can be applied to gain the asymptotic behavior of both the elastic and transition FFs but at low and intermediate energies, perturbative methods fail to converge due to the growth of the QCD coupling constant. Even though, on the theory side, significant progress in the past decades has been made \cite{Eichmann:2016yit,Ramalho:2023hqd}, a model-independent prediction for the TFFs at low energies $|q^2|< 1 \text{ GeV}^2$ which is dominated by the meson cloud contributions, is still lacking. To fill in this gap, based on dispersive formalism that includes the $\pi\pi$ rescattering effects in the s-channel and nucleon and $\Delta$ exchange in the t/u channel, we, for the first time, calculate the $N^*(1520)-N$ transition FFs in both space-like and time-like region for $|q^2|\leq 1 \text{GeV}^2$. On the experimental side, 
The space-like FFs have been measured by Jefferson lab (e.g.\cite{Aznauryan:2012ba}) and the time-like FFs have been extracted by the HADES experiment \cite{HADES:2022vus}, which offers interesting opportunities for a complete understanding of the low-energy TFFs. 
\section{Dispersive formalism}
\label{sec-dispersive formalism}
The $N^{*}(1520)$ has iso-spin $I=1/2$ and $J^{P}=3/2^{-}$. The electromagnetic matrix element is given by
 \begin{equation}
	\label{eq:def-transFF}
	\centering
	\langle N \vert j_\mu \vert N^* \rangle = 
	e \, \bar u_N(p_N) \, \Gamma_{\mu\nu}(q) \, u^\nu_{N^*}(p_{N^{*}})\,
	\end{equation}
	with
	\begin{equation}
\label{eq:def-TFF-Gam}
\begin{split}
	\Gamma^{\mu\nu}(q) := i \left(\gamma^\mu q^\nu - \slashed{q} g^{\mu\nu} \right) m_N F_1(q^2) 
	+ \sigma^{\mu\alpha} q_\alpha q^\nu F_2(q^2)+ i \left(q^\mu q^\nu - q^2 g^{\mu\nu} \right) F_3(q^2)   \,,
\end{split}
\end{equation}
where $q^{\mu}:=p_{N^{*}}^{\mu}-p_{N}^{\mu}$ and $F_{i}$ with $i=1,2,3$ are constraint-free FFs that are suitable for dispersive calculations. The dispersive method has been used by Uppsala Group \cite{Junker:2019vvy,Leupold:2017ngs, Granados:2017cib} to study the nucleon and hyperon FFs  and in this work we focus on the $N^*(1520)$ which is iso-vector dominant at low energies \cite{ParticleDataGroup:2022pth}. Diagrammatically, the FFs $F_i$, which are fully non-perturbative objects are represented by the grey blob in Eq.\,(\ref{eq lowE disp}). At low energies, the grey blob can be approximated by the right-hand diagram of Eq.\,(\ref{eq lowE disp}). We saturate the grey blob by a pion-pair because from the vector-meson dominance point of view, the $\rho$ meson couples strongly to the isovector TFF.
\begin{equation}
	\begin{split}
		\vcenter{\hbox{\begin{tikzpicture}[scale=0.5, transform shape]
					\begin{feynman}[every blob={/tikz/fill=gray!30,/tikz/inner sep=2pt}]
						\vertex (a) {$N^{*}$};
						\vertex [below right=2.5 cm of a, blob](b) {};
						\vertex [right=2.5 cm of b] (c) {$\gamma$};
						\vertex [below left=2.5 cm of b](d) {$N$};
						\diagram* {
							(a)-- [double,double distance=0.1ex,thick,with arrow=0.5,arrow size=0.15em](b) -- [photon] (c) 
							,
							(b) -- [ fermion] (d) 
							,
						};
					\end{feynman}
		\end{tikzpicture}}}
		&\approx\vcenter{\hbox{\begin{tikzpicture}[scale=0.5, transform shape]
					\begin{feynman}[every blob={/tikz/fill=red!30,/tikz/inner sep=2pt}]
						\vertex (a) {$N^{*}$};
						\vertex [below right=2.5 cm of a, blob](b) {BM};
						\vertex [right=2.5 cm of b, blob](c) {$F_v$};
						\vertex [right=2.5 cm of c] (d) {$\gamma$};
						\vertex [below left=2.5 cm of b](e) {$N$};
						\diagram* {
							(a)-- [double,double distance=0.1ex,thick,with arrow=0.5,arrow size=0.15em](b)--[scalar,half left](c)-- [scalar,half left](b),
							(c)--[photon](d),
							(b)--[fermion](e),};
					\end{feynman}
		\end{tikzpicture}}}+...=
    \includegraphics[valign = c,width=0.3\textwidth]{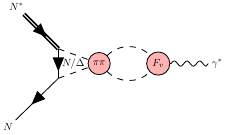}+...
	\end{split}
	\label{eq lowE disp}
\end{equation}
The red blob BM contains the $N^*(1520)\bar{N}\to2\pi$ \textbf{b}aryonic and \textbf{m}esonic interactions and the FF can be further approximated by the second equality in Eq.\,(\ref{eq lowE disp}) where the nucleon and $\Delta$ exchange in the crossing channels are explicitly shown. The $\Delta$ exchange has been shown to be important in FF calculations both in phenomenology and in the large $N_c$ limit \cite{Leupold:2017ngs}.
 The $F_v$ blob represents the pion-vector FF given by \cite{Leupold:2017ngs}
\begin{equation}
	\label{eq:FV-Omnes-alphaV}
	F_v(s) = (1+\alpha_V \, s) \, \Omega(s) \,
\end{equation}
with $\alpha_V= 0.12 \,\text{GeV}^{-2}$ and the Omnes function is $\Omega(s)$ defined as
\begin{equation}
	\Omega(s) = \exp\left\{ s \, \int\limits_{4m_\pi^2}^\infty \frac{\text{d}s'}{\pi} \, \frac{\delta(s')}{s' \, (s'-s-i \epsilon)} \right\}\,,
	\label{eq:omnesele}  
\end{equation}
where $\delta(s)$ is the p-wave pion phase shift.
As already stressed, the dispersion relations are formulated for the constraint-free FFs \cite{Junker:2019vvy}: 
\begin{equation}
F_i(q^2) = \frac{1}{12\pi} \, \int\limits_{4 m_\pi^2}^\infty \frac{ds}{\pi} \, 
		\frac{T_i(s) \, p_{\rm cm}^3(s) F_v^{*}(s)}{s^{1/2} \, (s-q^2-i \epsilon)}  +  F^{\rm anom}_i(q^2) + \ldots
\label{eq:dispbasicunsubtr}  
\end{equation}
where $T_i(s)\sim \bra{2\pi}\ket{N^{*}\Bar{N}}$ represents the hadronic p-wave partial wave amplitudes which are calculated using the following equation
\begin{eqnarray}
	T_i(s)  =  K_i(s) + \Omega(s) \, P_{i} + T_i^{\rm anom}(s) +\Omega(s) \, s \, 
	\int\limits_{4m_\pi^2}^\infty \, \frac{\text{d}s'}{\pi} \, 
	\frac{K_i(s') \, \sin\delta(s')}{\vert\Omega(s')\vert \, (s'-s-i \epsilon) \, s'} \,, \phantom{m} \,
	\label{eq:tmandel}
\end{eqnarray}
where $K_i$ are the tree-level diagrams as input. $P_{i=1,2,3}$ are subtraction constants which are essentially fit parameters. They parametrize the short-distance physics that is not explicitly included in our low-energy theory. The subtraction constants $P_i$ are fitted by matching to the S and D-wave decay widths $N^*(1520)\to N\rho$ \cite{ParticleDataGroup:2022pth}. The peculiar contribution  $F^{\rm anom}_i(q^2)$ in Eq.\,(\ref{eq:dispbasicunsubtr}) and $T_i^{\rm anom}(s)$ in Eq.\,(\ref{eq:tmandel}) original from the anomalous cut in the first Riemann sheet. This anomalous cut rooted mathematically in the Landau equations has been discussed recently (e.g. see \cite{Junker:2019vvy} and references therein).
\section{Preliminary results and outlook}
\begin{figure}[h!]
\begin{subfigure}{0.3\textwidth}
		\centering
		\includegraphics[width=\textwidth]{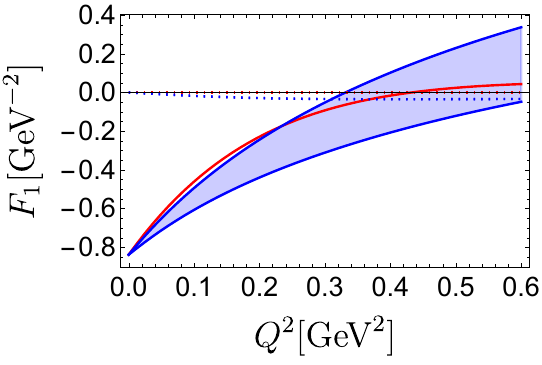}
	\end{subfigure}
	\begin{subfigure}{0.3\textwidth}
		\includegraphics[width=\textwidth]{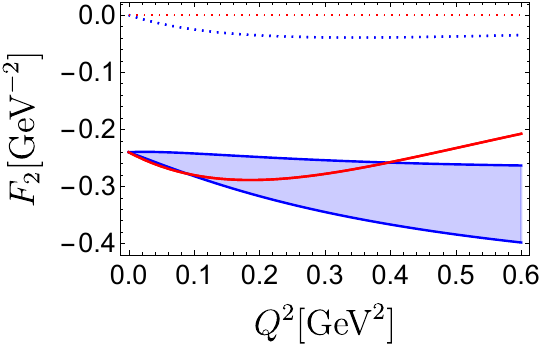}
	\end{subfigure}
	\begin{subfigure}{0.3\textwidth}
		\centering
		\includegraphics[width=\textwidth]{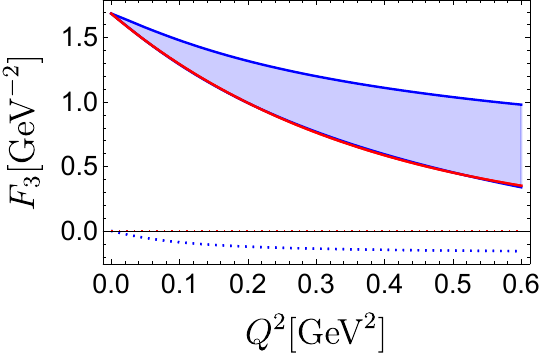}
	\end{subfigure}
		\caption{Red: Model dependent parametrization for isovector TFFs. We use Lisbon parametrization \cite{Eichmann:2018ytt} for proton, MAID parametrization for neutron. Blue: this work (preliminary). Full lines: real part, Dashed lines: imaginary part.}
	\label{fig spacelike}
\end{figure}

\begin{figure}[h!]
	\centering
	\begin{subfigure}{0.32\textwidth}
		\includegraphics[width=\textwidth]{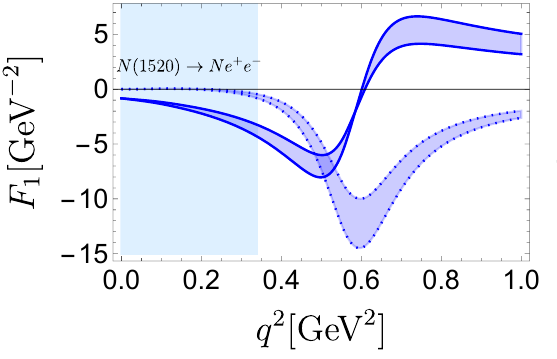}
	\end{subfigure}
	\begin{subfigure}{0.32\textwidth}
		\includegraphics[width=\textwidth]{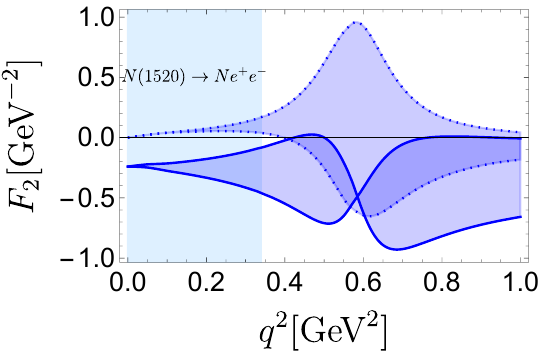}
	\end{subfigure}
	\begin{subfigure}{0.32\textwidth}
		\centering
		\includegraphics[width=\textwidth]{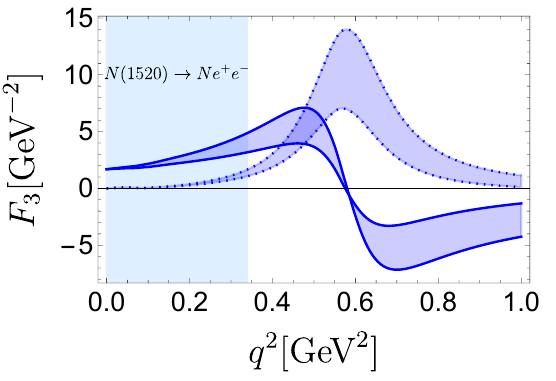}
	\end{subfigure}
	\caption{Time-like transition FFs (preliminary). Full lines: real part, dashed lines: imaginary part. The experimentally accessible region is defined as $q^2\in [4m_e^2,(m_N-m_{N^{*}})^2]$ which corresponds to the shaded-area.  }
	\label{fig timelike ff}
\end{figure}

\begin{figure}[h!]
	\centering
	\begin{subfigure}{0.45\textwidth}
		\includegraphics[width=\textwidth]{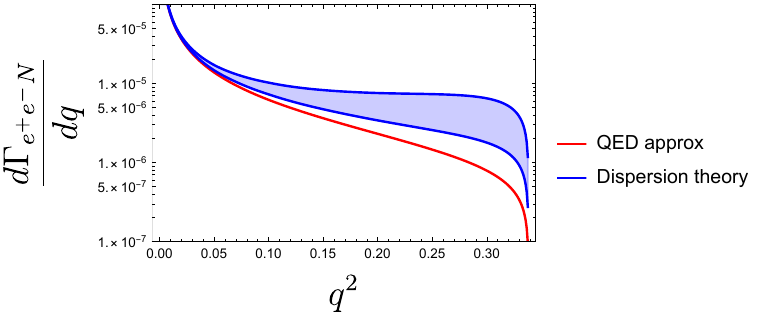}
	\end{subfigure}
\begin{subfigure}{0.45\textwidth}
	\includegraphics[width=\textwidth]{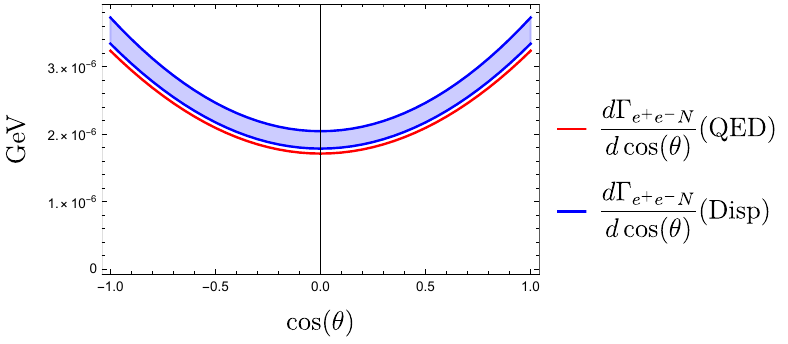}
\end{subfigure}
	\begin{subfigure}{0.45\textwidth}
		\includegraphics[width=\textwidth]{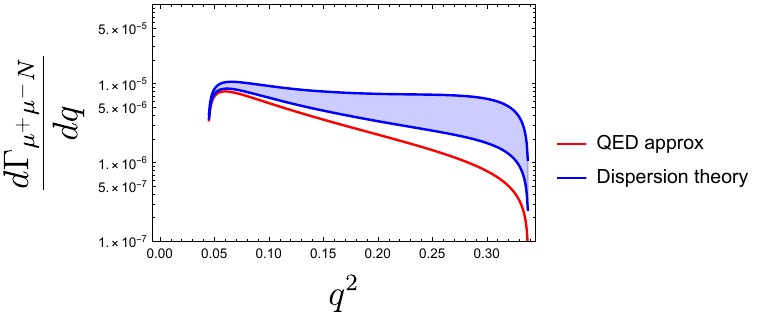}
	\end{subfigure}
	\begin{subfigure}{0.45\textwidth}
		\includegraphics[width=\textwidth]{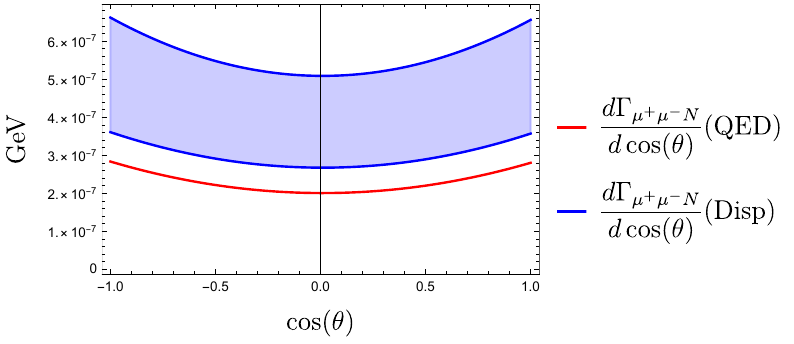}
  \end{subfigure}
	\caption{Predictions for electronic and muonic differential decay width (preliminary)}
	\label{fig electro decay}
 \end{figure}

The isovector FFs are defined as $\frac{1}{2}(F_i^{\text{proton}}-F_i^{\text{neutron}})$. 
 Based on the fitted subtraction constants $P_{1,2,3}$ introduced in Eq.\,(\ref{eq:tmandel}), space-like FFs are calculated as shown in  Fig.\,\ref{fig spacelike} where the experimental parametrization are also presented. Our preliminary result shows nice agreement with the experimental parametrization for $0<Q^2\leq 0.4\, \text{GeV}^2$. However, due to the fact that the MAID neutron parametrization has relatively large systematic uncertainty \cite{Tiator:2009mt}, we want to stress\footnote{ Especially, for $F_2$ at low energies the data parametrization shows that there is a broad dip which is probably due to systematic errors. Future measurements on the neutron TFFs will help to accurately extract the iso-vector TFFs and clarify the issue. } that the red curves should only be taken with a grain of salt.
 Time-like FFs are shown in Fig.\,\ref{fig timelike ff}. 
From the above predictions, differential decay widths $\frac{\Gamma_{N^{*}\to N l^+ l^-}}{dq}$ are shown in Fig.\,\ref{fig electro decay}. The preliminary results for Dalitz decay widths are  $\Gamma_{Nee}=4.9\pm0.3\, \text{keV}$, $\Gamma_{N\mu\mu}=0.85 \pm 0.25\,\text{keV}$. These numbers are larger than the QED estimates $\Gamma_{Nee}^{\rm QED}=4.4\, \text{keV}$, $\Gamma_{N\mu\mu}^{\rm QED}=0.4\,\text{keV}$, due to the influence of $\rho$ meson. These predictions can be tested by future experiments such as HADES and CBM \cite{Kiseleva:2011zz}.   In our dispersive framework, the only quantitative inputs are the S and D-wave $\Gamma_{N\rho}$ decay widths together with the hadronic two-body decay widths from $N^*(1520)\to N\pi,\Delta \pi$. Our dispersive calculations will certainly benefit from a more precise measurement of those quantities in the future.

\bibliography{./ref.bib}
\end{document}